\documentclass[preprint,12pt]{elsarticle}

\usepackage[utf8]{inputenc}
\usepackage[T1]{fontenc}

\usepackage{amsmath,amssymb}  
\usepackage{graphicx}
\usepackage{subfigure}
\usepackage{multirow}
\usepackage{booktabs}         
\usepackage{longtable}        
\usepackage{rotating}
\usepackage{float}
\usepackage{adjustbox}
\usepackage{url}
\usepackage{hyperref}  
\usepackage{makecell}
\usepackage{caption}    
\usepackage{soul}

\newcommand{\vecS}{\mathbf{s}}
\newcommand{\vecT}{\boldsymbol{\theta}}

\journal{Ecological Informatics}

\begin{document}

\begin{frontmatter}

\title{Foundation for unbiased cross-validation of spatio-temporal models for Species Distribution Modeling}

\author[1]{Diana Koldasbayeva\corref{cor1}}
\ead{diana.koldasbayeva@skoltech.ru}

\author[1]{Alexey Zaytsev}

\cortext[cor1]{Corresponding author}

\affiliation[1]{
  organization = {Skolkovo Institute of Science and Technology}, 
  addressline  = {Bolshoy Boulevard 30, bld. 1}, 
  city         = {Moscow}, 
  postcode     = {121205}, 
  country      = {Russia}
}

\begin{abstract}





Evaluating the predictive performance of species distribution models (SDMs) under realistic deployment scenarios requires careful handling of spatial and temporal dependencies in the data. Cross-validation (CV) is the standard approach for model evaluation, but its design can strongly influence the validity of performance estimates. When SDMs are intended for spatial or temporal transfer, random CV can lead to overoptimistic performance estimates due to spatial autocorrelation (SAC) among neighboring observations.

We benchmarked four machine learning algorithms (GBM, XGBoost, LightGBM, Random Forest) on two real-world presence–absence datasets --- a temperate plant and an anadromous fish --- under multiple CV designs: random, spatial, spatio-temporal, environmental, and forward-chaining. We evaluated two training-data usage strategies (LAST FOLD and RETRAIN) and applied extensive hyperparameter tuning within each CV scheme. Model skill was assessed on independent, out-of-time test sets using AUC, MAE, and correlation metrics.

Random splits overstated AUC by up to 0.16 and yielded mean absolute errors (MAE) 70–80\% higher than spatially blocked alternatives. Blocking at the empirical SAC range largely mitigated this bias. Blocking at the empirical SAC range mitigated this bias across datasets. Training-data usage influenced evaluation outcomes: LAST FOLD tended to yield smaller validation–test discrepancies in SAC-prone settings, whereas RETRAIN provided higher test AUCs in cases with weaker SAC. Boosted ensembles generally performed best across spatially blocked CV designs for both datasets.

We recommend a robust SDM workflow: (1) estimate SAC and construct spatial blocks accordingly; (2) tune hyperparameters using blocked cross-validation; (3) evaluate final models on external, out-of-time data. This pipeline enhances reliability for ecological forecasting under spatial and temporal shifts.

\end{abstract}

\begin{highlights}
\item Random CV overestimated AUC by up to 0.16 and raised MAE by $\approx 75 \%$.
\item Spatially blocked CV with tuning improved accuracy and ranking stability.

\item LAST FOLD reduced bias under SAC, while RETRAIN maximized data use.

\item Our pipeline enhances SDM reliability under spatial–temporal shifts.


\end{highlights}



\end{frontmatter}

\section{Introduction}
\label{sec:intro}

Species Distribution Modeling (SDM) is widely employed in ecology to understand biodiversity patterns and predict shifts in species ranges under changing environmental conditions. However, the validity of SDM predictions is based on careful handling of spatial autocorrelation (SAC), the phenomenon by which observations located closer together in space tend to exhibit greater similarity than those further apart~\cite{tobler1970computer, lennon2000red, dormann2007effects, ord1995local}. 
SAC violates the assumption of independent observations in statistical analyses, potentially introducing biased parameter estimates, overoptimistic model performance, and misleading ecological inferences~\cite{getis2009spatial, negret2020effects}. 
As such, the mitigation of SAC in SDM has received increasing attention, given its critical implications for conservation planning, species management, and climate change studies~\cite{ploton2020spatial, koldasbayeva2024challenges}.

Cross-validation (CV) is widely recognized as a pivotal tool for evaluating predictive models, including those used in SDM~\cite{stone1974cross, geisser1975predictive, roberts2017cross, hijmans2012cross}. 
Traditional CV  $K$-fold partitions the data into $K$ subsets (folds) randomly, iteratively using one subset for model testing while training on the remaining folds. Despite its popularity, this approach often overlooks SAC by randomly splitting data, thus allowing spatially correlated points to appear in both the training and test sets. 
This overlap leads to inflated estimates of model accuracy, as it does not measure a model’s ability to extrapolate to new, spatially distinct regions~\cite{dale2014spatial, roberts2017cross, valavi2018blockcv}.

To address these limitations, spatial CV strategies explicitly partition data into spatially independent folds~\cite{brenning2005spatial, roberts2017cross, valavi2018blockcv, ploton2020spatial}. 
For instance, geographical blocking creates folds separated by a minimum distance that ideally matches or exceeds the autocorrelation range~\cite{roberts2017cross}. 
Environmental blocking clusters locations based on feature similarity rather than pure geographic distance, ensuring that training and test sets encompass distinct ranges of predictor variables~\cite{valavi2018blockcv, roberts2017cross}. 
These methods yield more conservative, but arguably more realistic, estimates of predictive performance in spatially structured data. However, optimal blocking distances or buffer sizes can be difficult to determine, and environmental blocking may overlook key geographic constraints, limiting its applicability in long-term forecasts under changing climate conditions~\cite{santini2021assessing}.

While spatial CV techniques represent a significant step forward, several gaps persist. 
First, SAC is rarely considered alongside temporal dependencies. 
Given the growing usage of SDMs to predict range shifts under climate change, it is critical to evaluate how well these models generalize across both space and time~\cite{svenning2011applications, elith2009species}. 
Incorporating temporal dimensions into blocking strategies can help avoid overly optimistic performance estimates when projecting into future scenarios or unmeasured time periods.

The second gap relates to model deployment after spatial CV. Although spatial CV is commonly used to evaluate model performance~\cite{roberts2017cross, ploton2020spatial}, it is not clear how the final models should be trained and applied. In classical machine learning (ML) workflows, CV is primarily used for model selection, followed by retraining on the full dataset using the chosen hyperparameters. However, retraining after spatial CV can reintroduce SAC, potentially undermining the separation that the CV scheme was designed to enforce. This raises the question: should we instead use the model trained on the final fold of CV (hereafter referred to as "LAST FOLD"), which preserves strict spatial independence but relies on less data?

In our formulation, LAST FOLD refers to a strategy where the final model is not retrained on the full dataset, but instead taken directly from the last training fold used in spatial CV. This approach ensures that the training and test sets remain spatially separated, consistent with the original CV partitioning. However, a practical consequence is that only a subset of the data --- typically one spatial block --- is used for model deployment. 
Although rarely formalized in ecological modeling, the LAST FOLD strategy reflects broader principles of out-of-sample validation—such as those used in time-series forecasting via rolling-origin evaluation~\cite{hyndman2018forecasting}. Similarly, ecological studies emphasizing environmental or geographic novelty between training and test sets~\cite{hirzel2006evaluating, mila2022nearest} have implicitly followed related ideas. However, the explicit trade-offs introduced by LAST FOLD --- particularly between spatial independence and reduced training size --- remain underexplored in the SDM literature and merit further investigation.

Although both LAST FOLD and RETRAIN preserve formal separation between training and test data, RETRAIN can accidentally reintroduce SAC-related leakage by using the entire dataset, including spatially proximal observations to the test set. In contrast, LAST FOLD ensures strict geographic separation between training and test regions, at the cost of excluding potentially informative nearby data. This highlights the subtle but important difference between nominal independence (via splitting) and effective independence (via spatial distance), which is central to assessing generalization in spatially autocorrelated settings.

In parallel, spatial CV not only raises deployment questions but also increases the computational burden of model fitting. While spatial CV already increases the computational cost of model training by requiring multiple model fits in spatially disjoint folds, many studies nonetheless avoid additional expense by relying on default model settings, for example, using Random Forest with a fixed number of trees~\cite{roberts2017cross, ploton2020spatial, wadoux2021spatial} - without systematic hyperparameter tuning. In contrast, in our study we performed extensive hyperparameter optimization within each CV design, accepting the added computational burden in order to obtain fair and reliable performance estimates.
However, failure to optimize hyperparameters can substantially affect predictive performance and model selection, as shown in general ML contexts~\cite{probst2019tunability, probst2019hyperparameters}, and thus may obscure the true capabilities of spatial CV workflows.

We evaluated several CV schemes — random, spatial (SP), spatio-temporal (SPT), environmental (ENV) blocking, and TimeSeriesSplit (TSS) — applied across four ML algorithms (Random Forest, Gradient Boosting, XGBoost, and LightGBM).

Each CV design was used exclusively for hyperparameter tuning and model evaluation. For every algorithm --- CV combination, we tested 100 random hyperparameter configurations and scored model performance by area under the ROC curve on validation folds and an out-of-time test set, with the main goal of identifying CV strategies that best approximate out-of-time performance.

After hyperparameter tuning, we compared two alternative final-model training strategies:
(i) \emph{LAST FOLD}, which reuses the model fitted in the last training fold generated by the CV splitter, thereby preserving strict spatial independence but using fewer data; and
(ii) \emph{RETRAIN}, which refits the model on the full dataset using the tuned hyperparameters, maximizing data use but potentially reintroducing SAC.

We used two real-world presence-absence datasets - \emph{Gentianella campestris} (a temperate plant) and \emph{Thaleichthys pacificus} (an anadromous fish) — with contrasting spatial and temporal characteristics, allowing us to test how SAC and dataset dynamics affect validation reliability and transferability.
We find that naïve validation can lead to substantial errors not only in AUC but also in mean absolute error (MAE) and correlation metrics, highlighting that the impact of SAC extends across multiple dimensions of model evaluation.

\textbf{Our contributions are threefold:}
\begin{itemize}
    \item We provide the first empirical comparison of two model deployment strategies --- \emph{LAST FOLD} and \emph{RETRAIN} --- highlighting the trade-off between maintaining spatial independence and maximizing training data.

    \item We conduct a systematic evaluation of spatial and spatio-temporal CV strategies across diverse ecological contexts, identifying when blocking aligned with SAC range improves validation reliability.

    \item We demonstrate that performing hyperparameter tuning within spatial CV designs significantly enhances both predictive accuracy and validation robustness, emphasizing the need for joint tuning and proper validation in ecological modeling.
\end{itemize}

Together, these elements support a practical and conservative SDM workflow suited for applications where spatial or temporal transferability is essential.

\begin{figure}[!h]
    \centering
    \includegraphics[width=1.0\textwidth]{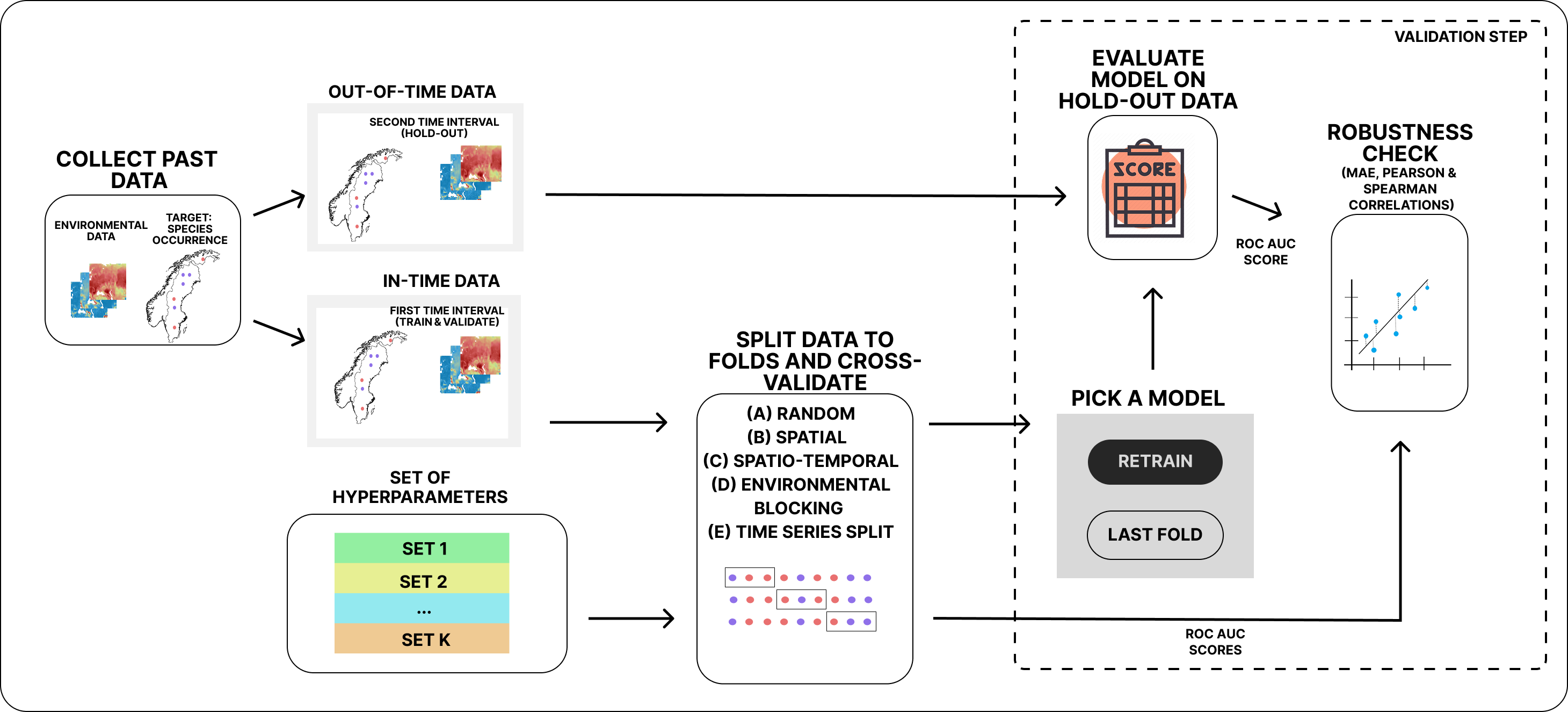}
    \caption{Overview of our modeling and validation design.
Past environmental and occurrence data are split into two time intervals: one for model training and validation, and another held out for final testing. Within the training interval, we apply several CV strategies - random, spatial, spatio-temporal, environmental blocking and TimeSeriesSplit - each tested with a range of hyperparameter sets and spatial distances to evaluate model performance using ROC AUC. The best-performing model is selected using either the RETRAIN or LAST FOLD strategy and then evaluated on the temporally independent hold-out data. To assess robustness, we compare AUC scores from CV and hold-out evaluation using mean absolute error (MAE), Pearson correlation ($r$), and Spearman correlation ($\rho$).}
    \label{fig:worklflow}
\end{figure}

\section{Methods}
\label{sec:Methods}

\subsection{Pipeline}
To construct a robust predictive model, we followed a systematic pipeline that included data collection, preprocessing, model selection, spatial CV, hyperparameter tuning, and performance evaluation (Figure \ref{fig:worklflow}).

Given our focus on spatio-temporal validation and CV design, it is essential to use species datasets that span different time intervals and exhibit contrasting spatial and temporal structures. To this end, we selected two complementary test cases:
(1) \emph{Gentianella campestris}
(2) \emph{Thaleichthys pacificus}.


To ensure temporal independence, the data were split into two time periods: in-time and out-of-time. Details of species  are provided in Section~\ref{subsec: species}.
This temporal split framework allows us to evaluate model performance under realistic forecasting scenarios where models must generalize to new time periods, while also testing whether spatial CV methods provide robust performance estimates even when the ultimate validation involves temporal rather than purely spatial independence.

To forecast species distributions, we selected ensemble-based models, including Random Forest and Gradient Boosting, with advanced variations of the latter, like XGBoost and LightGBM, due to their consistently strong performance in SDM applications~\cite{pichler2023machine}. These models were trained and validated using a range of spatial CV strategies, incorporating spatial, environmental, and a novel spatio-temporal blocking methods featuring varying distances. The full details of these methods are provided in Section~\ref{subsec: Blocking}.

We employed a systematic random search algorithm to identify optimal model configurations for each CV strategy, evaluating 100 random hyperparameter configurations for each of the four models across nine cross-validation schemes. Details on the hyperparameter search and selected configurations are presented in Section~\ref{subsec: models}.

To assess model performance, we used error and correlation-based metrics (details in Section \ref{subsec: evaluation}).

\label{subsec: pipeline}

\subsection{Data}
\label{subsec: species}
Two species were analyzed.
The first, \textit{Gentianella campestris}, a temperate plant with relatively stable distributions over time, for which we compiled a new dataset from Global Biodiversity Information Facility (GBIF)~\cite{gbif_2022} and national sources, details are described in the following Section~\ref{subsubsec:gentianella}. The second is \emph{Thaleichthys pacificus}, a culturally and ecologically important smelt of the north Pacific~\cite{bradburn2011}. We used a ready dataset from this work, and the details about it in Section~\ref{subsubsec:eulachon}.

\subsubsection{Gentianella campestris}
\label{subsubsec:gentianella}
\paragraph{Study area}
The study area spans from 31$^\circ$10' to 71$^\circ$10' East longitude and from 3$^\circ$55' to 55$^\circ$20' North latitude, including regions within Norway and Sweden. 
It has $843215.4$ $\mathrm{km}^2$ area.

This region features a diverse climate, including temperate maritime conditions in the south and subarctic to Arctic climates in the north. Prominent geographic elements comprise extensive coastlines along the Baltic Sea and North Sea, as well as mountainous terrain in the western parts~\cite{ketzler2021climate}.

\paragraph{Occurrence data}
Due to a lack of temporally-resolved studies on SDM, we selected the species \emph{Gentianella campestris}, a small herbaceous biennial flowering plant from the \emph{Gentianaceae} family, native to Europe. 
We collected occurrence data from GBIF database~\cite{gbif_2022}, which aggregates publicly available biodiversity records, including contributions from citizen science platforms, national monitoring programs, and herbarium collections. 

We downloaded 25381 occurrence records of \emph{Gentianella campestris}, restricted to the territories of Sweden and Norway and spanning the years 1960–2022. The dataset combines information from 55 sources. We used both presence and absence records. Records not identified to the species level or lacking coordinates were excluded. To mitigate spatial clustering and sampling bias, we applied partial spatial thinning using the \texttt{spThin} R package~\cite{aiello2015spthin}, setting a minimum distance of 500 meters between retained points.
This distance was chosen pragmatically: larger thresholds (e.g., 1 km) led to severe reductions in occurrences of the minority class (presence), potentially compromising model performance. Thus, 500 m represented a balance between reducing spatial sampling bias and retaining sufficient sample size, consistent with previous findings that thinning reduces spatial bias~\cite{BORIA201473} but excessive thinning can harm model accuracy when detections are rare~\cite{steen2021spatial}.
After preprocessing, the total number of absence records is 2959, and presence is 3168.

We note that this thinning distance is finer than the resolution of our climatic predictors (2.5 arc-minutes). Consequently, multiple thinned occurrence records can fall within a single climate grid cell. While this means these records are not independent in terms of the broad-scale climate values, the 500 m thinning was essential to reduce the strong SAC present in the raw data at local scales.

The spatial distribution of presence and absence locations is shown in Figure S1 (Supporting Information). The data from 2003–2018 were used for modelling, as this interval provided sufficient temporal coverage with available absences.

\paragraph{Environmental predictors}

We used 26 environmental predictors, including bioclimatic variables, soil properties, and elevation. We did not perform feature selection due to our chosen models, which inherently possess the ability to deal with high-dimensional data and manage a mix of informative and less relevant features.

We used \texttt{raster}~\cite{hijmans2015package}, \texttt{rgdal}~\cite{bivand2015package}, \texttt{sf}~\cite{pebesma2018simple}, \texttt{terra}~\cite{hijmans2022package} R packages to prepare environmental data for modeling. The study region, defined using merged shapefiles, was used to mask and crop environmental rasters. All layers were standardized to the WGS84 CRS and stacked into a single dataset. Table S1 (Supporting Information) lists the variables used in this study, and we refer the reader to the Worldclim~\cite{fick2017worldclim} and SoilGrids~\cite{hengl2017soilgrids250m} projects for detailed descriptions.

\subparagraph{Climate Data Processing}
To create temporally dynamic predictors for our spatio-temporal models, we utilized annual climate data. We downloaded yearly data for minimum temperature, maximum temperature, and total precipitation for the period 1984–2018 from the WorldClim database at a 2.5 arc-minute resolution. 

We processed this data in two distinct ways to create different sets of predictors:
\begin{enumerate}
    \item For random, spatial, environmental CV, we created two long-term averages representing distinct historical periods: \textbf{1984–2002} and \textbf{2003–2018}. These averages were converted into bioclimatic variables using the \texttt{dismo}~\cite{hijmans2017package} R package.
    \item For the spatio-temporal cross-validation (SPT-CV) models, we created four overlapping temporal intervals to match our blocking structure: \textbf{2003–2006}, \textbf{2007–2010}, \textbf{2011–2014}, and \textbf{2015–2018}. The annual climate data was averaged within each of these 4-year intervals before being converted to bioclimatic variables. This resulted in a unique set of climatic predictors for each spatio-temporal fold, allowing the model to incorporate temporal climate trends.
\end{enumerate}

\subparagraph{Static Predictors}
Soil properties (silt, sand, coarse fragments, bulk density, and soil organic carbon at 5–15 cm depth) were obtained from the SoilGrids database. Elevation data was sourced from WorldClim. These variables were treated as static in time over the study period.
As climatic variables were obtained at 2.5 arc-minute resolution, soil and elevation layers were resampled to this same grid for consistency.

\subsubsection{Thaleichthys pacificus}
\label{subsubsec:eulachon}
We used the U.S.\ West–Coast ground-fish bottom-trawl surveys~\cite{bradburn2011}. After filtering to hauls that recorded eulachon (\textit{Thaleichthys pacificus}) or a confirmed absence, the working data set contains 3\,181 geo-referenced trawl records collected between 2003--2012 (10 years). The spatial extent spans $41.0^{\circ}\mathrm{N}$--$48.5^{\circ}\mathrm{N}$ and $125.9^{\circ}\mathrm{W}$--$124.0^{\circ}\mathrm{W}$ (Washington--Oregon shelf).

From the original dataset, we selected the following environmental predictors: sea bottom temperature (Gtemp\_c), sampling depth (BEST\_DEPTH\_M), and swept area (AREA\_SWEPT\_MSQ). Hauls with missing values for these covariates were excluded. These covariates were not originally at a uniform spatial resolution. To support our spatial analysis and create a continuous prediction surface, we rasterized them across the study region at a 0.25$^\circ$  resolution. This resolution was chosen to match the methodology of the foundational spatiotemporal distribution model for eulachon \cite{ward2015using}, which itself used the native grid of Sea Surface Temperature product~\cite{reynolds2007daily}, a standard dataset in marine ecology provided at this spatial scale.

To evaluate temporal generalization, we split the data into a training set (2006--2012) and a temporal test set (2003--2005). This split allocates the more data-rich recent period to training to maximize the model's information, while using the earlier, less-data-rich period for testing. Spatial coordinates and year were retained for spatio-temporal blocking but were excluded from the feature matrix for model fitting, consistent with our approach for \emph{Gentianella campestris}.

The original dataset was highly imbalanced, with presence points (occurrences of \textit{Thaleichthys pacificus}) comprising approximately 9\% of the total observations. To improve model learning in this setting, we applied Synthetic Minority Over-sampling Technique (SMOTE) within each cross-validation fold.
We chose SMOTE over alternative methods for two key reasons: first, to avoid discarding a large portion of the already limited dataset and the valuable ecological information contained in the absence records; and second, to generate new examples in the feature space and help the model learn a more robust decision boundary for the rare presence class.

We used a conservative target ratio of 30\% presence after oversampling. This approach retained all original absence samples and synthetically generated additional presence records to achieve the desired balance. On average, this resulted in approximately 3 to 5 times more synthetic presences than original ones per fold, depending on the actual class distribution in the training split.

\subsection{Model training workflow}
\label{subsec: val strategies}

\subsubsection{Cross-validation for the identification of the best hyperparameters}

The RETRAIN strategy involves a two-step process aimed at optimizing the use of available data for model training. Initially, the training dataset \( D = \{(\mathbf{X}_i, y_i)\}_{i=1}^N \) is partitioned into \( k \)-folds such that
\[
D = \bigcup_{j=1}^k D_j, \quad D_j \cap D_{j'} = \emptyset \quad \text{for } j \neq j'.
\]
Here, \( \mathbf{X}_i \) denotes the \textit{feature vector} for the \( i \)-th observation, and \( y_i \) is the corresponding \textit{target variable} (a scalar, e.g., species presence or absence).

For each fold \( j \), a model \( M_{\vecT} \) with hyperparameters $\vecT$ is trained on the training subset \( D \setminus D_j \) and validated on the testing subset \( D_j \). The optimal hyperparameters \( \vecT^* \) are determined by minimizing the average loss across all folds:
\[
\vecT^* = \arg\min_{\vecT} \frac{1}{k} \sum_{j = 1}^k L(M_{\vecT}^{(j)}, D_j),
\]
where \( M_{\vecT}^{(j)} \) denotes the model with hyperparameters $\vecT$ trained on \( D \setminus D_j \), and \( L(M, D_j) \) represents the loss of model \( M \) evaluated on dataset \( D_j \). We used one minus the ROC AUC (Receiver Operating Characteristic Area Under the Curve) score as the loss function for this binary classification task.

This procedure, while standard, introduces a potential selection bias because the test performance is conditioned on the hyperparameter set chosen by its highest CV score (see Supporting Information S5 for a theoretical discussion).

Once the optimal hyperparameters ($\theta^*$) were identified through cross-validation, two distinct final-model training strategies were applied: RETRAIN and LAST FOLD, described below.

\subsubsection{RETRAIN}
Once the best hyperparameters \( \vecT^* \) are identified, the model is retrained on the entire dataset \( D \) using these parameters to produce the final RETRAIN model:
\[
M_{\mathrm{final}} = M_{\vecT^*}(D).
\]
We note that $\vecT^*$ can be different from the hyperparameters that provide the best error for a holdout test sample, but we hope that they are as close as possible.

This approach ensures that all available data contribute to the final model, potentially enhancing its robustness and predictive accuracy.

However, a notable limitation of the RETRAIN strategy is its disregard for spatial or temporal autocorrelation. By combining training and testing subsets during the retraining process, this method may lead to optimistic bias, particularly in datasets with strong SAC, where dependencies between data points inflate performance metrics. Nevertheless, RETRAIN remains a practical choice for scenarios where maximizing data utilization is prioritized over strict adherence to spatial or temporal independence.

\subsubsection{LASTFOLD}

Instead of retraining the model on the entire dataset, as in the RETRAIN strategy, the LAST FOLD method trains the final model \( M_{\mathrm{final}} \) using only the training subset of the last fold \( D_k \) and the optimized hyperparameters:
\[
M_{\mathrm{final}} = M_{\theta^*}(D_k).
\]

This approach ensures that the final model reflects the spatial and temporal structure of the data, preserving dependencies within the last fold. This is important because RETRAIN, by combining all folds, may reintroduce SAC leakage: neighboring points from previously held-out folds can bias the model and inflate its apparent performance.

In contrast, LAST FOLD trains the final model on a single spatial block from the CV process, ensuring that the training data are spatially independent of the held-out test set. Because the training and test folds are drawn from similar spatial–temporal partitions, the resulting data distributions may more closely resemble those encountered in sliding-window or forward-chaining evaluation schemes. This property can be advantageous in scenarios where spatial and temporal autocorrelation strongly influence model performance, as it reduces the mismatch between validation and test conditions.
However, the method sacrifices some training data, as only a portion of the dataset is used for constructing the final model. This trade-off must be carefully considered when choosing the LAST FOLD strategy.

\subsection{Methods of cross-validation}
\label{subsec: Blocking}

\paragraph{Random cross-validation}
Random k-fold CV involves uniformly random splitting of a dataset into $k$ subsets (folds) and using $k - 1$ folds for training and the remaining one for testing in each iteration. We repeated this process multiple times, and the performance metrics were averaged to evaluate the model's generalization ability. 
Random CV was performed using R package \texttt{caret}~\cite{kuhn2020package}. 
For \emph{Gentianella campestris}, we used k = 5 folds, and for \emph{Thaleichthys pacificus}, k = 3 folds, consistent with our other CV approaches.

\paragraph{Spatial blocking}
\label{paragrpaph: spatial cv}
Spatial CV is a technique designed to account for spatial dependencies in data by dividing the study area into geographically distinct blocks~\cite{pohjankukka2017estimating}. 

To implement spatial blocking, the study area is divided into non-overlapping spatial blocks. For \emph{Gentianella campestris}, we used 5 spatial folds, and for \emph{Thaleichthys pacificus}, 3 spatial folds, reflecting the different data densities of the two species.  The size of the block determines the spatial separation scale between the training and validation sets. While points near block boundaries may still be in close proximity, larger block sizes increase the average distance between training and validation samples and better reduce the influence of fine-scale SAC.~\cite{roberts2017cross}.

The optimal block size for spatial CV was determined using the \texttt{blockCV} R package~\cite{valavi2018blockcv}, specifically its \texttt{cv\_spatial\_autocor} function. This function automatically fits variograms to each continuous raster variable in the dataset to estimate the effective range of SAC. The SAC range represents the distance at which observations become spatially independent, based on the variogram analysis.

In our study, the SAC range for \emph{Gentianella campestris} was calculated as approximately 422 km. This distance was used as the optimal block size for spatial CV, ensuring minimal SAC between training and validation blocks. To explore the effect of varying block sizes on model performance, we also tested smaller (200 km) and larger (600 km) block sizes. These blocks are referred to as \textbf{SP 200}, \textbf{SP 422}, and \textbf{SP 600}, respectively. 

For \emph{Thaleichthys pacificus}, the SAC range was estimated at approximately 85 km based on the environmental variogram. Accordingly, we used \textbf{SP 40}, \textbf{SP 85}, and \textbf{SP 120} blocks to examine the sensitivity of model performance to block size.

Table~\ref{tab:cvtypes} lists all CV schemes used in the experiments.

\begin{table}
\centering
\caption{Cross-validation (CV) types and their descriptions for both species.}
\label{tab:cvtypes}
\resizebox{\textwidth}{!}{%
\begin{tabular}{llcc}
\toprule
\textbf{Species} & \textbf{CV Type} & \textbf{Description} & \textbf{Key details} \\
\midrule
\multirow{8}{*}{\makecell{\emph{Gentianella} \\ \emph{campestris}}} 
& \textbf{Random}  & Random splitting & Cross-validation with random partitions, ignoring spatial structure. \\
& \textbf{SP 200}  & Spatial blocking (200 km) & Spatial blocks with a distance threshold of 200 km to separate folds. \\
& \textbf{SP 422}  & Spatial blocking (422 km) & Spatial blocks with a distance threshold of 422 km, aligned with SAC range. \\
& \textbf{SP 600}  & Spatial blocking (600 km) & Spatial blocks with a distance threshold of 600 km to separate folds. \\
& \textbf{ENV}     & Environmental blocking & Clustering data based on environmental similarity using K-means. \\
& \textbf{SPT 200} & Spatio-temporal blocking (200 km) & Spatial blocks (200 km) combined with temporal folds (4 years). \\
& \textbf{SPT 422} & Spatio-temporal blocking (422 km) & Spatial blocks (422 km) combined with temporal folds (4 years). \\
& \textbf{SPT 600} & Spatio-temporal blocking (600 km) & Spatial blocks (600 km) combined with temporal folds (4 years). \\
& \textbf{TSS} & TimeSeriesSplit  & Training on all past temporal blocks and validation on the next block (forward-chaining). \\
\midrule
\multirow{8}{*}{\emph{Thaleichthys pacificus}} 
& \textbf{Random}  & - & - \\
& \textbf{SP 40}   & Spatial blocking (40 km) & Spatial blocks with a distance threshold of 40 km to separate folds. \\
& \textbf{SP 85}   & Spatial blocking (85 km) & Spatial blocks with a distance threshold of 85 km, aligned with SAC range. \\
& \textbf{SP 120}  & Spatial blocking (120 km) & Spatial blocks with a distance threshold of 120 km to separate folds. \\
& \textbf{ENV}     & - & - \\
& \textbf{SPT 40}  & Spatio-temporal blocking (40 km) & Spatial blocks (40 km) combined with temporal folds (2-3 years). \\
& \textbf{SPT 85}  & Spatio-temporal blocking (85 km) & Spatial blocks (85 km) combined with temporal folds (2-3 years). \\
& \textbf{SPT 120} & Spatio-temporal blocking (120 km) & Spatial blocks (120 km) combined with temporal folds (2-3 years). \\
& \textbf{TSS} & TimeSeriesSplit  & Training on all past temporal blocks and validation on the next block (forward-chaining).\\
\bottomrule
\end{tabular}
}
\end{table}

\paragraph{Environmental blocking}
The objective of this method is to create clusters of data points that are homogeneous in terms of their environmental conditions while ensuring that each cluster contains a balanced representation of classes in the target variable.
We used k = 5 environmental clusters for \emph{Gentianella campestris} and k = 3 for \emph{Thaleichthys pacificus}.
This balancing was uniquely necessary for environmental blocking because clustering in feature space can create partitions with severe class imbalance, whereas spatial and temporal folds naturally maintain the original class distribution of their respective regions or time periods.
Environmental blocking is based on the K-mean clustering method. All features, including climate, soil, and elevation properties, were split into clusters. 
We chose the number of clusters by the elbow method~\cite{kodinariya2013review}. 
The proposed algorithm considered that each cluster must include both classes. We named this method \textbf{ENV}.

\paragraph{Spatio-temporal blocking}

In adopting a spatio-temporal (SPT) CV strategy, our primary goal was to account for both the temporal dynamics and spatial structure of the data. Initially, we considered assigning each year to a separate fold; however, due to data limitations and the relatively small year-to-year variability in climate predictors, we grouped years into broader intervals. For \emph{Gentianella campestris}, the data was partitioned into four temporal intervals: 2003–2006, 2007–2010, 2011–2014, and 2015–2018.

Within each temporal interval, we applied spatial CV by dividing the data into five spatial folds (blocks). When one fold was used as the validation set, the remaining folds from the same temporal interval served as training data, ensuring that no records from other time periods were used for model fitting. This design prevented spatial leakage across time while maintaining independence between training and validation sets. Each spatial fold was used in turn as the validation set, with the remaining folds as training data. This resulted in 20 iterations (5 spatial folds × 4 time intervals), yielding a comprehensive assessment of model performance across spatio-temporal partitions. This method allowed us to evaluate both spatial and temporal generalization of the models.

In this research, we refer to this methodology as \textbf{SPT 200}, \textbf{SPT 422}, and \textbf{SPT 600}, corresponding to 200 km, 422 km, and 600 km temporal block sizes, respectively.

For \emph{Thaleichthys pacificus}, we applied the same spatio-temporal blocking framework, adapted to the characteristics of the dataset. The data was divided into three temporal intervals: 2005–2008, 2009–2010, and 2011–2012. Due to the smaller number of observations available in certain years, spatial blocking within each interval used 3 spatial folds, with spatial block sizes of \textbf{SPT 40}, \textbf{SPT 85}, and \textbf{SPT 120}, aligned with the species' SAC range.

\paragraph{TimeSeriesSplit}

Time Series Split (\textbf{TSS}) CV, also known as forward-chaining, is a standard approach for temporally ordered data. It ensures models are trained only on past observations and validated on future ones, thereby avoiding future-to-past leakage.

For \emph{Gentianella campestris}, we used $T = 4$ temporal blocks, and for \emph{Thaleichthys pacificus}, $T = 3$ temporal blocks, corresponding to the temporal intervals defined in the spatio-temporal blocking section.

The dataset is divided into $T$ contiguous blocks
\[
D = \bigcup_{t=1}^{T} D_t, \qquad \mathrm{Time}(D_1) < \cdots < \mathrm{Time}(D_T).
\]

At fold $j \in \{1,\dots,T-1\}$, training uses all past blocks and validation uses the next block:
\[
D_{\mathrm{train}, j} = \bigcup_{t=1}^{j} D_t, 
\qquad 
D_{\mathrm{val}, j} = D_{j+1}.
\]

For each hyperparameter vector $\boldsymbol{\theta}$, the model $M_{\boldsymbol{\theta}}$ is fitted on $D_{\mathrm{train}, j}$ and evaluated on $D_{\mathrm{val}, j}$, yielding
\[
\mathrm{AUC}_{\mathrm{mean}}(\boldsymbol{\theta}) 
= \frac{1}{T-1} \sum_{j=1}^{T-1} 
\mathrm{AUC}\!\left(M_{\boldsymbol{\theta}}(D_{\mathrm{train}, j}),\, D_{\mathrm{val}, j}\right).
\]

The optimal configuration is then selected as
\[
\boldsymbol{\theta}^* 
= \arg\max_{\boldsymbol{\theta}} \; \mathrm{AUC}_{\mathrm{mean}}(\boldsymbol{\theta}).
\]

Finally, the model $M_{\mathrm{final}}$ is trained either by RETRAIN on all in-time data or by LAST FOLD using only the most recent training window, and in both cases is evaluated on the independent out-of-time test set $D_{\mathrm{test}}$.

The temporal splitting scheme for both datasets is described in detail in Supporting Information S3.


\subsection{Models}
\label{subsec: models}

Classical ML algorithms tend to be used for tasks such as SDM with subsequent applications for identifying conservable or restorable areas~\cite{beery2021species}. 
Random Forest and Boosting Decision Tree are some of the most popular tools in these research studies\cite{pichler2023machine}. 
Using the R packages \texttt{gbm}~\cite{greenwell2019package}  , \texttt{randomForest}~\cite{liaw2002classification}, \texttt{xgboost}~\cite{chen2019package} and \texttt{lightgbm}~\cite{lightgbm}, we adopted these models to predict the distribution of \emph{Gentianella campestris} and \emph{Thaleichthys pacificus}. All models were trained for binary classification of presence versus absence.
Gradient Boosting (GBM), XGBoost, and LightGBM were optimized with binary log-loss (cross-entropy), while Random Forest used Gini impurity as the splitting criterion and produced probabilistic predictions.

We utilized a random search for hyperparameter tuning, selecting 100 hyperparameter sets at random from predefined ranges for each model. Table S2 lists the hyperparameters for each model. 

\paragraph{Random Forest}
Random Forest is an ensemble learning method that constructs multiple decision trees during training and combines their predictions to produce a more robust and accurate model~\cite{breiman2001random}. 
It introduces randomness in the tree-building process by selecting random subsets of features and data points, reducing overfitting and improving generalization. 

\paragraph{Gradient Boosting}
GBM is a powerful ensemble learning method that builds a predictive model by combining the predictions of multiple weak learners, typically decision trees~\cite{friedman2001greedy}. 
This method minimizes the loss function iteratively by adding new decision trees that focus on the mistakes made by the previous ensemble of trees. 
It sequentially fits new trees to the residuals of the previous predictions, gradually reducing prediction errors.


\paragraph{XGBoost} 
One of the variations of GBM models is XGBoost, which is widely used in various ML competitions and applications. Like other GBM methods, XGBoost employs decision trees as weak learners and focuses on minimizing the loss function iteratively~\cite{chen2016xgboost}. However, XGBoost stands out due to its unique features, such as a regularized objective function for improved model generalization, which helps prevent overfitting. Additionally, XGBoost includes enhancements like efficient handling of missing values, parallel processing for faster training, and advanced regularization techniques. The final prediction combines the individual forecastings from all the sequentially added decision trees.

\paragraph{LightGBM}
LightGBM is another variation of GBM models that creates a predictive model through a collection of weak learners, often decision trees. Like other GBM methods, LightGBM iteratively adds decision trees that focus on correcting the errors of previous trees, improving prediction accuracy as the boosting process continues~\cite{ke2017lightgbm}. However, LightGBM stands out due to its unique approach to building trees, using a histogram-based learning algorithm for efficient training on large datasets and faster execution. By splitting the data into histograms, LightGBM significantly accelerates the algorithm, making it especially effective for large datasets.

\subsection{Evaluation metrics}

Having split our data into two temporal segments, namely, in-time and out-of-time data — we trained the models using the in-time data. During training, CV was applied to in-time data to tune hyperparameters and assess model performance. At this stage, we record the ROC AUC scores for each fold, as it is a widely used metric to evaluate the performance of binary classification.

After tuning the hyperparameters, we validated the models on the out-of-time data and recorded the ROC AUC scores for this independent test set. 
The mean ROC AUC scores from the cross-validation (averaged across all folds) and the out-of-time validation were then compared to assess the consistency of the model’s performance across these phases.

To quantify the alignment between CV and out-of-time results, we computed the MAE, Pearson's correlation, and Spearman's correlation coefficients between the ROC AUC scores. The role of these metrics is detailed below:
\label{subsec: evaluation}

\subsubsection{Model evaluation}
\paragraph{ROC AUC}
ROC AUC evaluates the model's ability to distinguish between presence and absence during both CV and out-of-time testing. It measures the area under the ROC curve, which plots the true positive rate (sensitivity) against the false positive rate (1-specificity) at various threshold values. A higher ROC AUC score signifies better model discrimination, with a score of $1$ indicating perfect discrimination and $0.5$ representing random guessing~\cite{bradley1997use, shabani2018assessing}. 

\subsubsection{Validation robustness}
In this subsection, we introduce quality metrics suitable for evaluating the quality of scores obtained using a specific CV strategy. 
We take into account that we are interested not only in a specific predicted quality value, but as well in ranking models according to obtained scores.
So, we assume that we have model quality scores $\hat{\vecS} = \{ \hat{s}_i \}_{i = 1}^m$ obtained from a validation method and true scores obtained using a large separate test sample $\vecS = \{s_i\}_{i = 1}^m$.
From these scores, we compute the vector of ranks $R(\vecS)$, where instead of specific values, we have their corresponding ranks at a specific place of a vector.

\paragraph{MAE}
MAE measures the absolute difference between the ROC AUC scores from CV and out-of-time testing. This metric quantifies the average absolute difference between predicted and actual values~\cite{willmott2005advantages}. A lower MAE indicates that the model's predictions are closer to the true values, while a higher MAE signifies greater prediction errors:
\[
\mathrm{MAE}(\hat{\vecS}, \vecS) = \frac{1}{m} \sum_{i=1}^{m} |s_i - \hat{s}_i|.
\]

\paragraph{Pearson Correlation}
Pearson correlation evaluates the linear relationship between the ROC AUC scores from the two phases. It calculates the strength and direction of the linear association between the scores. A Pearson correlation coefficient close to $1$ indicates a strong positive linear relationship, close to $-1$ suggests a strong negative linear relationship and near $0$ implies a weak or no linear relationship.
It has the following form:
\[
\rho_{\mathrm{Pearson}}(\hat{\vecS}, \vecS) = \frac{\sum_{i=1}^{m}(\hat{s}_i - \bar{\hat{s}})(s_i - \bar{s})}{\sqrt{\sum_{i = 1}^{m} (\hat{s}_i - \bar{\hat{s}})^2 \sum_{i = 1}^{m}(s_i - \bar{s})^2}},
\]
where $\bar{s} = \frac{1}{m} \sum_{i = 1}^m s_i$ is a mean value of the vector $\vecS$, and $\bar{\hat{s}} = \frac{1}{m} \sum_{i = 1}^m \hat{s}_i$.

\paragraph{Spearman Correlation}
Spearman Correlation, also known as Spearman's rank correlation coefficient, measures the monotonic relationship between the quality scores from CV and out-of-time testing. Unlike Pearson Correlation, Spearman does not assume a linear relationship and can capture nonlinear monotonic associations. It has the following form where there are no ties:
\[
\rho_{\mathrm{Spearman}}(\hat{\vecS}, \vecS) = \rho_{\mathrm{Pearson}}(R(\hat{\vecS}), R(\vecS)) = 1 - \frac{6 \sum_{i = 1}^m d_i^2}{m (m^2 - 1)},
\]
where $d_i = R(\hat{\vecS})_i - R(\vecS)_i$ is the difference between the two ranks of each observation.

This $\rho_{\mathrm{Spearman}}$ is particularly useful for evaluating consistency in rank-order performance, even in the presence of nonlinearity in the data.

\section{Results}

\subsection{Robustness Evaluation}
\label{sec:robustness}

\begin{figure}[!htbp]
  \centering
  \includegraphics[width=0.75\linewidth]{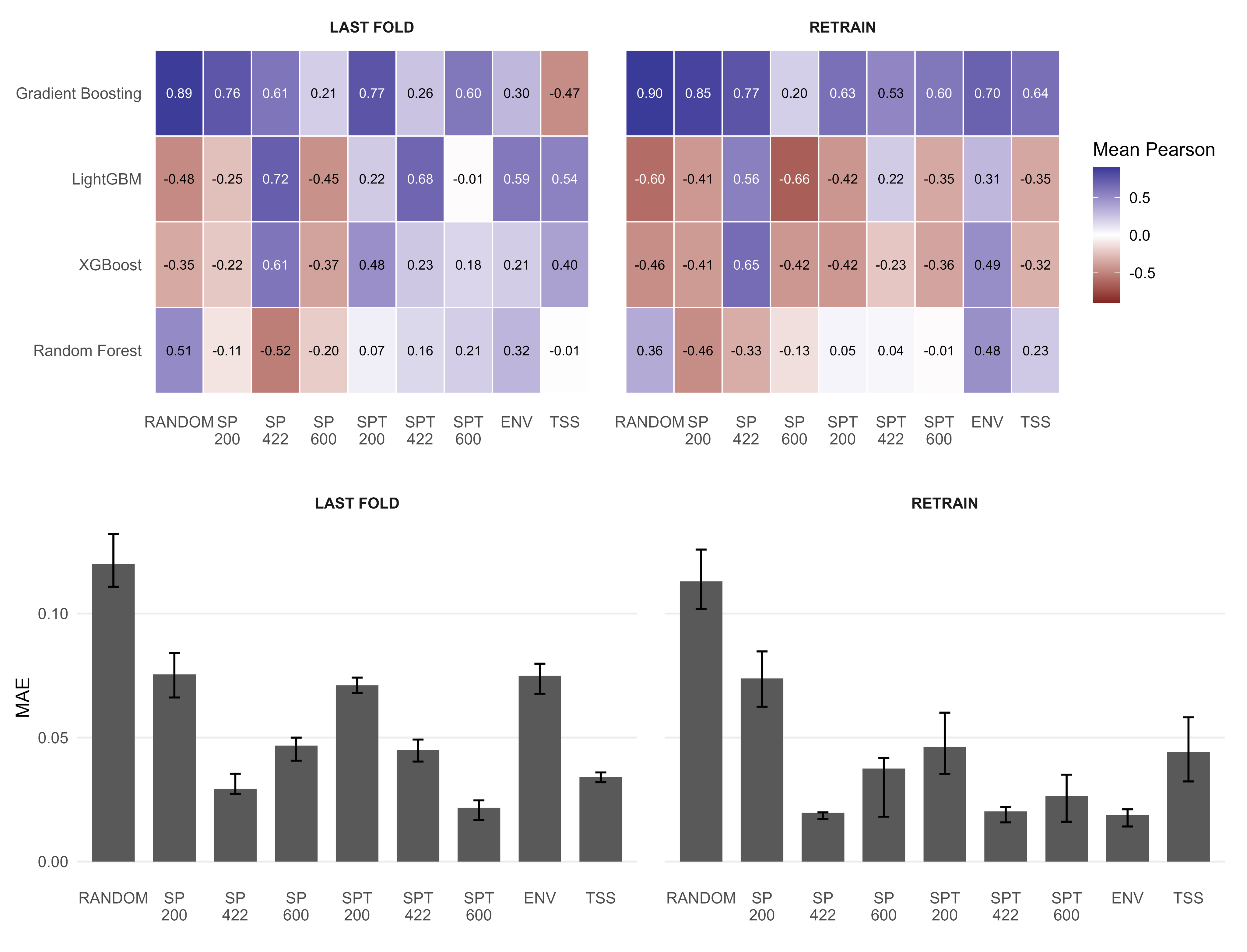}
  \caption*{(a) \emph{Gentianella campestris}}
\includegraphics[width=0.75\linewidth]{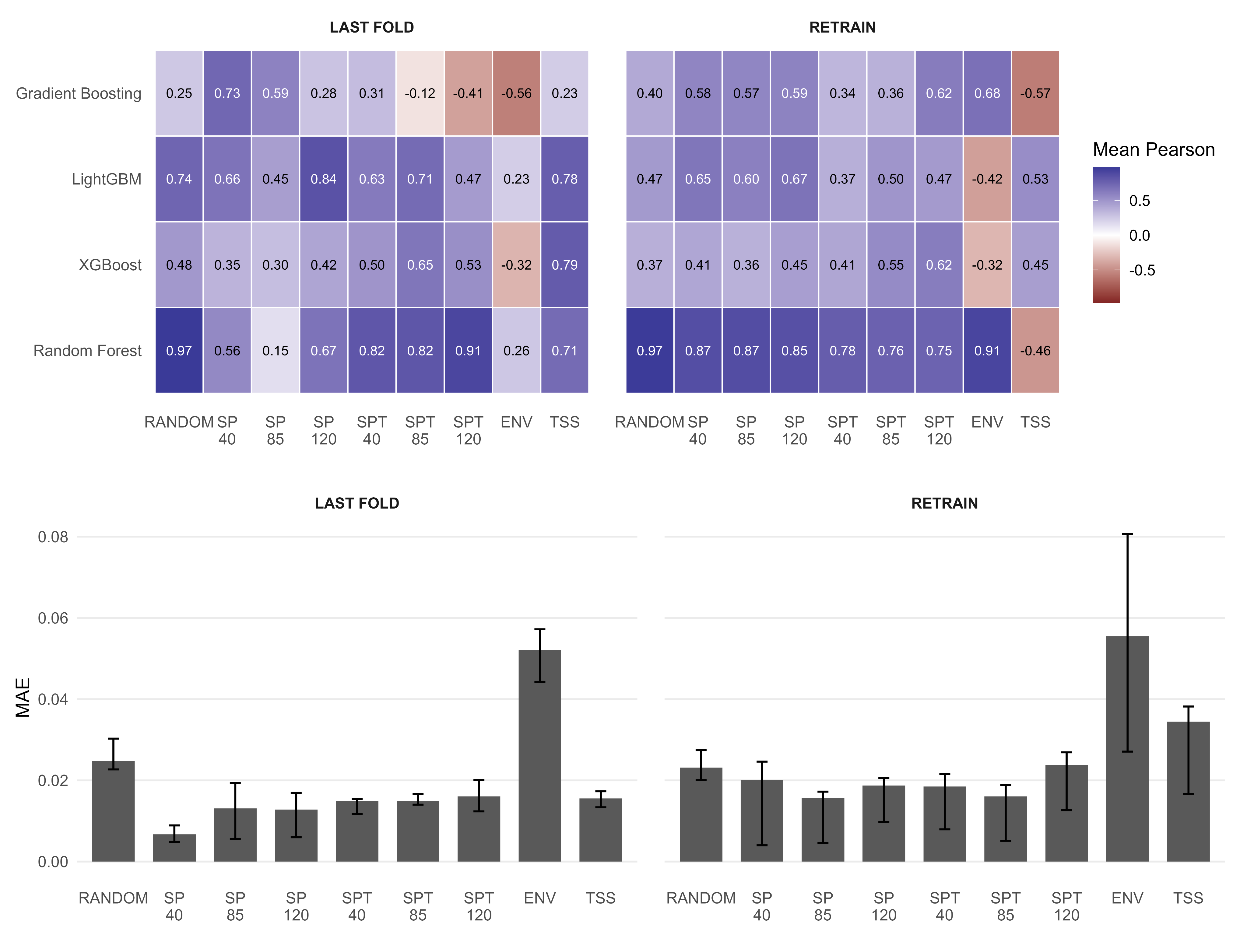}
  \caption*{(b) \emph{Thaleichthys pacificus}}
  \caption{
  Comparison of model performance under different CV schemes and training strategies.
  Each panel shows the mean Pearson correlation (top) and the mean MAE $\pm$ IQR (bottom),
  where IQR denotes the interquartile range (25th–75th percentile) across models.
  Spatial and spatio-temporal blocking produced lower errors and more consistent correlations
  than random or environmental splits.
  }
  \label{fig:mae_pearson}
\end{figure}

We evaluated how well different CV schemes predict out-of-time performance and compared two final training strategies (RETRAIN and LAST FOLD). 
Figure~\ref{fig:mae_pearson} summarizes the agreement between in-fold CV metrics and temporally independent test performance.
The upper panels show the \emph{mean} Pearson correlation (averaged across folds) by model and CV scheme, while the lower panels display the \emph{mean} MAE across models with IQR (25th–75th percentile), capturing between-model variability.
Full numeric results (including Spearman’s~$\rho$) are provided in Tables~S4.1–S4.2.

Across both datasets, Figure~\ref{fig:mae_pearson} demonstrates that spatial and spatio-temporal blocking aligned to the SAC range consistently improved validation robustness. Random and environmental CV schemes showed the largest discrepancies between CV and test metrics, often yielding unstable or overly optimistic correlations.

\subsubsection{Proper spatial blocking improves validation robustness}

For \emph{G. campestris}, ENV and SP~422 under RETRAIN achieved the lowest MAE (0.019–0.020) and the highest correlations (Spearman~0.619).  
Under LAST FOLD, spatial blocking at the SAC range (SP~422) also improved MAE (0.035 vs.~0.120 for Random CV).  
Overall, spatial blocking consistently outperformed Random CV and produced more reliable validation–test alignment.

For \emph{T. pacificus}, SP~40–85 (matching SAC $\approx$85~km) provided the best robustness.  
Random CV produced unstable results: while certain models (RF) showed strong correlations, others (XGB and LGB) exhibited inconsistent or negative correlations, and MAE was generally higher.

TSS generally underperformed spatial blocking for \emph{G.~campestris} (0.044~MAE vs.~0.019–0.038 for ENV/SPT), but achieved superior temporal generalization for \emph{T.~pacificus} (Spearman~$\rho=0.630$ vs.~$\rho<0.626$ in RETRAIN). This highlights its niche usefulness when temporal signals outweigh SAC.

\begin{figure}[!htbp]
  \centering
  \includegraphics[width=0.65\linewidth]{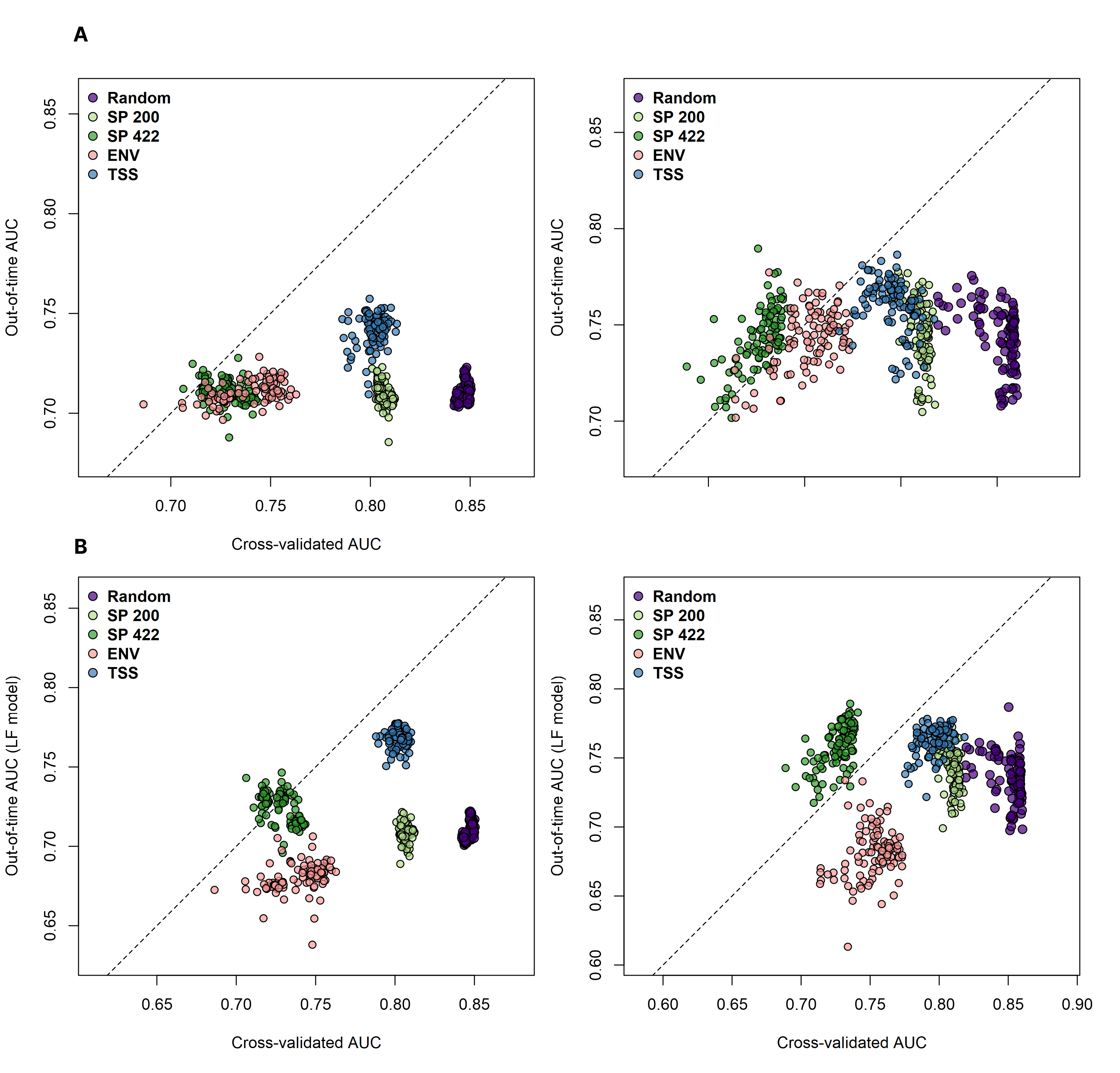}
  \caption*{(a) \emph{Gentianella campestris}}
  \includegraphics[width=0.65\linewidth]{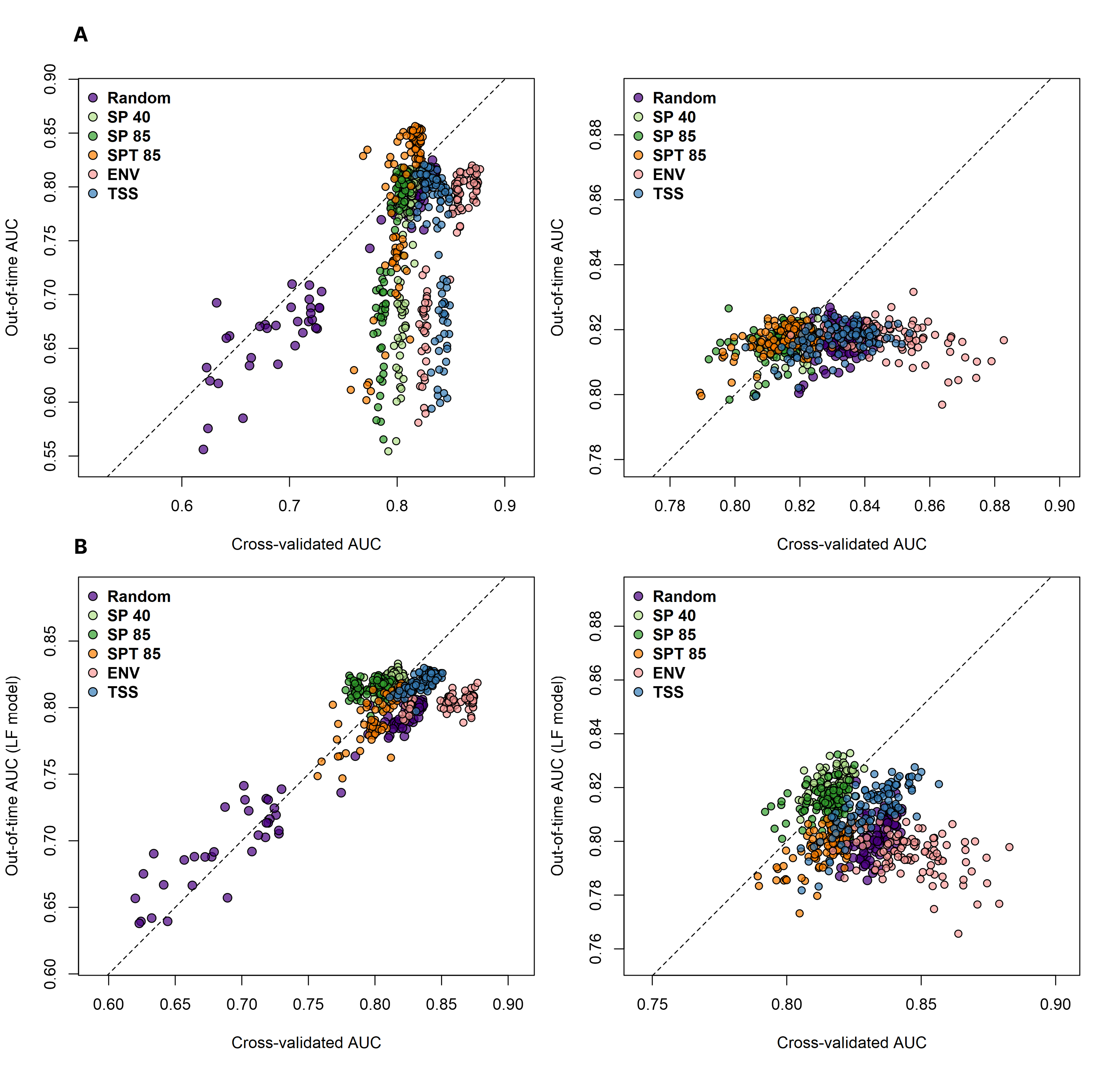}
  \caption*{(b) \emph{Thaleichthys pacificus}}
  \caption{
  Relationship between cross-validated and out-of-time AUC across CV schemes and training strategies.
Panels A–B show RETRAIN and LAST FOLD results, with XGBoost (left) and Random Forest (right).
Each point represents one hyperparameter configuration; the dashed diagonal marks perfect agreement between validation and test AUC.
  }
  \label{fig:scores_xgb_rf}
\end{figure}

Figure ~\ref{fig:scores_xgb_rf} illustrate these patterns: points under Random CV  consistently fall above the diagonal, confirming systematic overestimation of validation AUC.
By contrast, spatial and spatio-temporal blocking (SP, SPT) yield clusters closer to the diagonal, indicating better alignment between validation and test performance.
The comparison of RETRAIN (top) and LAST FOLD (bottom) further shows that LAST FOLD narrows the spread of points in SAC-dominated settings, reducing variance in overestimation.
Full scatterplots for all models and CV designs are provided in Figures S4.1–S4.2 (Supporting Information), showing consistent trends.

\subsubsection{LAST FOLD shows potential but requires caution}

We also assessed how restricting the final evaluation to the last fold (rather than retraining on all folds) affects alignment with out-of-time performance.  

Across datasets, the LAST FOLD results were mixed.  
For \emph{T.~pacificus}, LAST FOLD achieved lower MAE under SP~40 (0.007) than RETRAIN (0.016), indicating a closer agreement between validation and test performance.  
In contrast, for \emph{G.~campestris}, RETRAIN yielded better MAE (0.020) than LAST FOLD (0.029) under SP~422.  
Correlation metrics were typically higher under RETRAIN, especially for ENV CV (Spearman~$\rho=0.619$), suggesting more reliable hyperparameter ranking.

Overall, while LAST FOLD may mitigate overestimation in strongly clustered datasets, its advantages are not consistent across contexts. Because it restricts the training data and may lower model generalization, we do not recommend it as a default choice. Rather, we include it for comparability with previous spatial SDM studies and emphasize that robust, SAC-aware CV designs—particularly spatial or spatio-temporal blocking—remain the most reliable basis for model validation.

\subsection{Model evaluation}
\label{sec: model eval}

We next compared ML algorithms under different CV and training strategies.
Tables~\ref{tab:oracle_gentianella}-\ref{tab:oracle_eulachon}  report the \textit{oracle} ROC~AUC---i.e., the best test AUC among 100 hyperparameter combinations per model.
This analysis addresses a practical question: Which CV scheme and training strategy (LAST FOLD vs. RETRAIN) would a practitioner select if only the top-performing configuration were to be deployed?


\begin{table}[ht]
\centering
\caption{\textbf{Maximum AUC performance under the RETRAIN and LAST FOLD strategies for \emph{Gentianella campestris}.} Each cell reports the best AUC computed on the independent out-of-time test set across hyperparameter runs. \\
\textit{Note:} These results are conditioned on selecting the best hyperparameter set via cross-validation. This introduces a form of selection bias; comparisons should be interpreted with this in mind.}

\label{tab:oracle_gentianella}
\begin{tabular}{llccccc}
\toprule
\textbf{Strategy} & \textbf{CV} & \textbf{GBM} & \textbf{RF} & \textbf{XGB} & \textbf{LGB} & \textbf{Average} \\
\midrule
\multirow{9}{*}{RETRAIN} 
& Random    & 0.761 & 0.723 & 0.776 & 0.779 & 0.760 \\
& SP 200    & 0.761 & 0.723 & 0.777 & 0.779 & 0.760 \\
& SP 422    & 0.764 & 0.728 & \textbf{0.790} & 0.779 & 0.765\\
& SP 600    & 0.765 & 0.733 & 0.776 & 0.779 & 0.763 \\
& ENV       & 0.762 & 0.728 & 0.777 & 0.779 & 0.762 \\
& SPT 200   & 0.758 & 0.721 & 0.771 & 0.779 & 0.757 \\
& SPT 422   & 0.760 & 0.722 & 0.775 & 0.779 & 0.759 \\
& SPT 600   & 0.761 & 0.731 & 0.772 & 0.779 & 0.761 \\
& TSS       & 0.768 & 0.757 & 0.786 & 0.775 & \textbf{0.772} \\
\midrule
\multirow{9}{*}{LAST FOLD} 
& Random    & 0.750 & 0.722 & 0.787 & 0.773 & 0.758 \\
& SP 200    & 0.753 & 0.722 & 0.772 & 0.778 & 0.756 \\
& SP 422    & 0.777 & 0.746 & 0.789 & 0.783 & 0.774 \\
& SP 600    & 0.797 & 0.770 & \textbf{0.798} & 0.795 & \textbf{0.790} \\
& ENV       & 0.689 & 0.706 & 0.734 & 0.702 & 0.708 \\
& SPT 200   & 0.727 & 0.732 & 0.725 & 0.721 & 0.726 \\
& SPT 422   & 0.715 & 0.733 & 0.711 & 0.705 & 0.716 \\
& SPT 600   & 0.722 & 0.692 & 0.712 & 0.708 & 0.709 \\
& TSS       & 0.773 & 0.778 & 0.778 & 0.776 & 0.776 \\
\bottomrule
\end{tabular}
\end{table}

\begin{table}[ht]
\centering
\caption{\textbf{Maximum AUC performance under the RETRAIN and LAST FOLD strategies for \emph{Thaleichthys pacificus}.} Each cell reports the best AUC computed on the independent out-of-time test set across hyperparameter runs. \\
\textit{Note:} These results are conditioned on selecting the best hyperparameter set via cross-validation. This introduces a form of selection bias; comparisons should be interpreted with this in mind.}
\label{tab:oracle_eulachon}
\begin{tabular}{llccccc}
\toprule
\textbf{Strategy} & \textbf{CV} & \textbf{GBM} & \textbf{RF} & \textbf{XGB} & \textbf{LGB} & \textbf{Average} \\
\midrule
\multirow{9}{*}{RETRAIN} 
& Random & 0.826 & 0.825 & 0.827 & 0.817 & 0.824 \\
& SP 40  & 0.827 & 0.822 & 0.824 & 0.817 & 0.823 \\
& SP 85  & 0.826 & 0.818 & 0.827 & 0.817 & 0.822 \\
& SP 120 & 0.828 & 0.818 & 0.823 & 0.817 & 0.822 \\
& ENV    & 0.826 & 0.820 & 0.832 & 0.817 & 0.824 \\
& SPT 40 & 0.827 & 0.855 & 0.824 & 0.817 & 0.831 \\
& SPT 85 & 0.826 & \textbf{0.857} & 0.826 & 0.817 & \textbf{0.832} \\ 
& SPT 120& 0.826 & 0.855 & 0.825 & 0.817 & 0.831 \\
& TSS    & 0.830 & 0.821 & 0.826 & 0.817 & 0.824 \\
\midrule
\multirow{9}{*}{LAST FOLD} 
& Random & 0.820 & 0.808 & 0.822 & 0.820 & 0.818 \\
& SP 40  & 0.830 & 0.833 & 0.833 & 0.820 & \textbf{0.829} \\
& SP 85  & 0.826 & 0.824 & 0.833 & 0.822 & 0.826 \\
& SP 120 & 0.819 & 0.835 & 0.818 & 0.780 & 0.813 \\
& ENV    & 0.813 & 0.819 & 0.809 & 0.806 & 0.812 \\
& SPT 40 & 0.813 & 0.821 & 0.817 & 0.798 & 0.812 \\
& SPT 85 & 0.814 & 0.818 & 0.809 & 0.790 & 0.808 \\
& SPT 120& 0.808 & 0.821 & 0.810 & 0.787 & 0.807 \\
& TSS    & \textbf{0.836} & 0.830 & 0.828 & 0.817 & 0.828 \\
\bottomrule
\end{tabular}
\end{table}

For \emph{G.~campestris}, whose SAC range is $\approx$422~km, SP~422 was optimal under RETRAIN (average~0.765, XGB~0.790).  
Under LAST FOLD, slightly coarser SP~600 achieved the highest average AUC (0.790, XGB~0.798).  
Blocks smaller than the SAC range (SP~200) produced lower AUCs and weaker reliability metrics.  
Nevertheless, SP~422 under LAST FOLD remained competitive (AUC~0.774) while yielding the lowest MAE (Table~S4.1), suggesting it as a robust default.
However, SP 422 under LAST FOLD remains competitive (AUC 0.774) while offering the lowest MAE (Table~\ref{tab:oracle_gentianella}), making it a robust default choice.

For \emph{T.~pacificus}, RETRAIN achieved its best overall performance at SPT~85 (average~0.832).  
LAST FOLD with SP~40 performed comparably (average~0.829) but exhibited lower MAE (0.007 vs.~0.013), indicating stronger transferability.  
Thus, SAC-aligned block sizes balance both accuracy and robustness.

At appropriate block widths (SP~422/600 for \emph{G.~campestris}, SP~85 for \emph{T.~pacificus}), LAST FOLD achieves performance competitive with RETRAIN while minimizing SAC leakage. For plant species, LAST FOLD clearly surpasses RETRAIN. For \emph{Thaleichthys pacificus}, RETRAIN shows slightly higher oracle AUC at SPT 85, but Section~\ref{sec:robustness} shows that these gains are not consistently reflected in validation robustness metrics (MAE and correlation), suggesting that LAST FOLD provides a more conservative and reliable choice for transferability.

Across both datasets, boosted ensembles (XGB, LGB) generally rank among the top-performing models, particularly under spatial blocking schemes. For \emph{Thaleichthys pacificus}, Random Forest achieves the highest scores under spatio-temporal blocks (0.857 in RETRAIN, 0.835 in LAST FOLD), while boosted methods dominate on the plant dataset. GBM is consistently competitive, but shows greater sensitivity to block size.

\section{Discussion}

\subsection{Implications for SDM validation}

Our results reinforce that SAC inflates performance estimates when ignored~\cite{roberts2017cross, ploton2020spatial, pohjankukka2017estimating, mila2022nearest, koldasbayeva2024challenges}. We confirm that spatial blocking aligned to the SAC range is essential for obtaining robust and transferable SDMs.

We evaluated both RETRAIN (final model trained on all folds) and LAST FOLD (final model trained only on the last validation fold) as part of our pipeline. These strategies represent common choices in recent SDM literature. Although RETRAIN sometimes achieved higher correlation metrics, LAST FOLD yielded a lower MAE between validation and out-of-time test performance in the \emph{Thaleichthys pacificus} dataset, suggesting reduced overoptimism under high SAC conditions.

However, this pattern did not generalize: for \emph{Gentianella campestris}, RETRAIN showed better validation-test alignment. These results highlight that model selection strategies can affect perceived robustness, particularly under SAC-driven bias, and should be interpreted carefully. Importantly, LAST FOLD should not be viewed as a general solution to SAC leakage, but rather as a conservative benchmark for comparison with earlier studies.

Spatio-temporal blocking (SPT) showed limited benefit, likely due to coarse temporal intervals (4-year blocks) and weak climate change trends in our data. In such contexts, spatial blocking alone may suffice, though SPT could prove valuable in studies of abrupt environmental change.

\subsection{Model selection and hyperparameter tuning}
While Random Forest remains one of the most widely used models in SDM research due to its simplicity~\cite{pichler2023machine, roberts2017cross, ploton2020spatial, schratz2019hyperparameter, adhikari2023global, maglietta2023environmental}, our results indicate that boosted ensembles (GBM, XGBoost, LightGBM) outperformed Random Forest in the majority of settings, consistently achieving top ranks in both oracle AUC and validation robustness (Sections~\ref{sec: model eval}, \ref{sec:robustness}). An exception was observed in the spatio-temporal RETRAIN configuration, where Random Forest achieved the highest AUC (0.857). These findings reinforce that algorithm performance depends critically on CV design.

Hyperparameter tuning is a well-recognized driver of model performance in SDM, with extensive work on MaxEnt optimization~\cite{radosavljevic2014making, syfert2013effects} and recent cross-algorithm toolkits for RF and other learners~\cite{vignali2020sdmtune, velazco2022flexsdm}.
Building on this foundation, we used a random search strategy (100 combinations per model) within every CV design to quantify how tuning interacts with spatial and spatiotemporal partitioning. Figure \ref{fig:scores_xgb_rf} reveals substantial performance dispersion throughout the hyperparameter space, reinforcing the need for systematic, CV-aware optimization when comparing SDM workflows.

This further underscores that model comparisons must be interpreted within the CV framework used. Comparisons across different model selection strategies (for example, LAST FOLD vs. RETRAIN) may conflate robustness with training volume, and thus should be contextualized rather than generalized.

\subsection{Limitations and future directions}

Our study provides a comprehensive comparison of cross-validation strategies, but several limitations warrant consideration. First, while our inclusion of two ecologically distinct species---a terrestrial plant and a marine fish---offers insights across ecosystem types, it limits broad taxonomic generalizations. The performance of CV strategies may vary across species with different dispersal abilities, habitat specificities, and data characteristics. Future research should validate these patterns across a wider range of taxa to develop more universally applicable guidelines for SDM evaluation.

Second, both datasets span limited time periods (approximately two decades for \emph{Gentianella campestris} and one decade for \emph{Thaleichthys pacificus}) with only moderate climate change signals. As longer-term datasets become available, future studies could better evaluate the potential of spatio-temporal blocking, particularly for species or regions experiencing stronger environmental trends.

Third, our focus on tree-based classifiers with presence-absence data, while methodologically consistent, leaves open questions about how these findings extend to other modeling approaches. Expanding comparisons to include presence-only methods (such as MaxEnt and Poisson point-process models) would broaden the applicability of our conclusions.

Looking forward, we identify two promising directions for strengthening SDM validation research. To complement real-world case studies, virtual species experiments could provide controlled testing of how specific data characteristics (e.g., prevalence, SAC strength, sample size) influence CV performance under known truth conditions. Such simulations would help isolate individual factors that are confounded in the observational data. Additionally, as more dynamic datasets become available, further work should explore forward-chaining and other temporally-aware validation strategies for stronger inference in spatiotemporal contexts.

Despite these limitations, our results support a practical validation recipe: \textbf{(i)} estimate the SAC range, \textbf{(ii)} block data at that scale, and \textbf{(iii)} tune models with blocked CV. The open-source workflow we provide enables rapid replication on additional taxa and regions, facilitating the broader validation needed to establish robust SDM evaluation standards.

\subsection{Selection bias}  

In our experiments, the final test performance was reported for the hyperparameter configuration that achieved the highest CV score.
While this reflects a common practice in model deployment, it introduces a well-known form of selection bias: the test performance is conditioned on the model chosen by its best validation score and may therefore overestimate the expected generalization ability.
This bias affects both RETRAIN and LAST FOLD strategies.
Furthermore, because these two strategies differ in the amount of training data used, part of their performance difference may arise from model capacity rather than validation robustness.
We acknowledge this limitation and recommend future work to explore unbiased estimators --- such as nested CV or analytical corrections --- to better isolate transfer robustness effects.
Nonetheless, the CV strategies examined here are widely used in ecological modeling due to their simplicity and robustness, and we believe that explicitly highlighting their limitations contributes to improving current practice and understanding of their biases (see Supporting Information S5 for a theoretical analysis).

\subsection{Considerations on evaluation metrics}
Our study relied on AUC as the primary evaluation metric due to its widespread adoption in species distribution modeling and its practicality for comparing diverse modeling scenarios. However, we acknowledge the recognized limitations of AUC, particularly its sensitivity to the geographic extent of the study area and its inability to distinguish between different types of spatial errors \cite{lobo2008auc}. Recent methodological advances suggest that alternative metrics like the Continuous Boyce Index ~\cite{hirzel2006evaluating} and spatially explicit evaluation approaches \cite{bracho2024spatially} could provide valuable complementary insights, especially for assessing model calibration and handling sampling biases. Future research comparing cross-validation strategies could benefit from employing a suite of evaluation metrics to assess different aspects of model performance.

\section{Conclusion}

We found substantial validation errors across all metrics—AUC, MAE, and correlation, when SAC was not properly accounted for, confirming that overoptimism is a general problem rather than a metric-specific one.

This study confirms that spatial CV aligned with the species’ empirical SAC range is critical for evaluating SDMs under spatial and temporal transfer scenarios. Across both datasets, spatial blocking consistently reduced overestimation and improved the reliability of out-of-time performance estimates.

Comparing two model training strategies, LAST FOLD and RETRAIN, revealed that although RETRAIN occasionally produced stronger validation --- test correlations, LAST FOLD more effectively reduced overestimation—particularly in spatially autocorrelated datasets such as Thaleichthys pacificus. This indicates that LAST FOLD provides a more conservative alternative when minimizing information leakage between validation and final training is a priority.

The potential of spatio-temporal blocking was more limited, reflecting the relatively weak temporal signals in our datasets. We therefore recommend SPT primarily in cases with pronounced interannual trends or abrupt temporal shifts.

Our systematic hyperparameter tuning across algorithms further showed that algorithm performance depends strongly on the CV design. While boosted ensembles generally outperformed Random Forest, the magnitude and consistency of these differences varied across partitioning schemes.

Taken together, these findings provide a practical framework for SDM validation: estimate the SAC range, apply spatial blocking, and, when spatial leakage is a concern, consider conservative training strategies such as LAST FOLD.
However, we emphasize that LAST FOLD should not replace RETRAIN as the default, but rather serve as a complementary approach in sensitivity analyses assessing the robustness of validation results.
Finally, incorporate CV-aware hyperparameter tuning. This pipeline enhances robustness and supports more reliable model deployment in conservation and ecological forecasting.


\section{Code availability}

For the data, preprocessing and modelling details to reproduce the calculations, we refer the reader to the repository of the project \url{https://github.com/Disha0903/spBlock_cv}.




\section{Author contributions statement}
Diana Koldasbayeva: Conceptualization, Methodology, Investigation, Writing – review \& editing, Writing – original draft, Visualization, Validation, Software, Formal analysis, Data curation. Alexey Zaytsev: Supervision, Methodology, Investigation, Conceptualization, Writing – review \& editing.

\section{Funding}
The research was supported by the Russian Science Foundation grant No. 25-11-00355, \url{https://rscf.ru/project/25-11-00355/}

\section{Additional information}

\textbf{Competing interests}

The authors declare no competing financial interests.

\bibliographystyle{elsarticle-num}
\bibliography{sample}  

@article{tobler1970computer,
  title={A computer movie simulating urban growth in the Detroit region},
  author={Tobler, Waldo R},
  journal={Economic geography},
  volume={46},
  number={sup1},
  pages={234--240},
  year={1970},
  publisher={Taylor \& Francis}
}

@article{roberts2017cross,
  title={Cross-validation strategies for data with temporal, spatial, hierarchical, or phylogenetic structure},
  author={Roberts, David R and Bahn, Volker and Ciuti, Simone and Boyce, Mark S and Elith, Jane and Guillera-Arroita, Gurutzeta and Hauenstein, Severin and Lahoz-Monfort, Jos{\'e} J and Schr{\"o}der, Boris and Thuiller, Wilfried and others},
  journal={Ecography},
  volume={40},
  number={8},
  pages={913--929},
  year={2017},
  publisher={Wiley Online Library}
}

@article{elith2009species,
  title={Species distribution models: ecological explanation and prediction across space and time},
  author={Elith, Jane and Leathwick, John R},
  journal={Annual Review of Ecology, Evolution and Systematics},
  volume={40},
  number={1},
  pages={677--697},
  year={2009}
}

@article{fick2017worldclim,
  title={WorldClim 2: new 1-km spatial resolution climate surfaces for global land areas},
  author={Fick, Stephen E and Hijmans, Robert J},
  journal={International journal of climatology},
  volume={37},
  number={12},
  pages={4302--4315},
  year={2017},
  publisher={Wiley Online Library}
}

@book{dale2014spatial,
  title={Spatial analysis: a guide for ecologists},
  author={Dale, Mark RT and Fortin, Marie-Jos{\'e}e},
  year={2014},
  publisher={Cambridge University Press}
}

@article{schratz2019hyperparameter,
  title={Hyperparameter tuning and performance assessment of statistical and machine-learning algorithms using spatial data},
  author={Schratz, Patrick and Muenchow, Jannes and Iturritxa, Eugenia and Richter, Jakob and Brenning, Alexander},
  journal={Ecological Modelling},
  volume={406},
  pages={109--120},
  year={2019},
  publisher={Elsevier}
}

@article{ord1995local,
  title={Local spatial autocorrelation statistics: distributional issues and an application},
  author={Ord, J Keith and Getis, Arthur},
  journal={Geographical analysis},
  volume={27},
  number={4},
  pages={286--306},
  year={1995},
  publisher={Wiley Online Library}
}

@article{ploton2020spatial,
  title={Spatial validation reveals poor predictive performance of large-scale ecological mapping models},
  author={Ploton, Pierre and Mortier, Fr{\'e}d{\'e}ric and R{\'e}jou-M{\'e}chain, Maxime and Barbier, Nicolas and Picard, Nicolas and Rossi, Vivien and Dormann, Carsten and Cornu, Guillaume and Viennois, Ga{\"e}lle and Bayol, Nicolas and others},
  journal={Nature communications},
  volume={11},
  number={1},
  pages={4540},
  year={2020},
  publisher={Nature Publishing Group UK London}
}

@article{pohjankukka2017estimating,
  title={Estimating the prediction performance of spatial models via spatial k-fold cross validation},
  author={Pohjankukka, Jonne and Pahikkala, Tapio and Nevalainen, Paavo and Heikkonen, Jukka},
  journal={International Journal of Geographical Information Science},
  volume={31},
  number={10},
  pages={2001--2019},
  year={2017},
  publisher={Taylor \& Francis}
}

@article{valavi2018blockcv,
  title={blockCV: An R package for generating spatially or environmentally separated folds for k-fold cross-validation of species distribution models},
  author={Valavi, Roozbeh and Elith, Jane and Lahoz-Monfort, Jos{\'e} J and Guillera-Arroita, Gurutzeta},
  journal={Methods in Ecology and Evolution},
  volume={10},
  number={2},
  pages={225--232},
  year={2018},
  publisher={Wiley Online Library}
}

@article{stone1974cross,
  title={Cross-validatory choice and assessment of statistical predictions},
  author={Stone, Mervyn},
  journal={Journal of the Royal statistical society: Series B (Methodological)},
  volume={36},
  number={2},
  pages={111--133},
  year={1974},
  publisher={Wiley Online Library}
}

@article{wadoux2021spatial,
  title={Spatial cross-validation is not the right way to evaluate map accuracy},
  author={Wadoux, Alexandre MJ-C and Heuvelink, Gerard BM and De Bruin, Sytze and Brus, Dick J},
  journal={Ecological Modelling},
  volume={457},
  pages={109692},
  year={2021},
  publisher={Elsevier}
}

@article{hengl2017soilgrids250m,
  title={SoilGrids250m: Global gridded soil information based on machine learning},
  author={Hengl, Tomislav and Mendes de Jesus, Jorge and Heuvelink, Gerard BM and Ruiperez Gonzalez, Maria and Kilibarda, Milan and Blagoti{\'c}, Aleksandar and Shangguan, Wei and Wright, Marvin N and Geng, Xiaoyuan and Bauer-Marschallinger, Bernhard and others},
  journal={PLoS one},
  volume={12},
  number={2},
  pages={e0169748},
  year={2017},
  publisher={Public Library of Science San Francisco, CA USA}
}

@article{liaw2002classification,
  title={Classification and regression by randomForest},
  author={Liaw, Andy and Wiener, Matthew and others},
  journal={R news},
  volume={2},
  number={3},
  pages={18--22},
  year={2002}
}

@article{chen2019package,
  title={Package ‘xgboost’},
  author={Chen, Tianqi and He, Tong and Benesty, Michael and Khotilovich, Vadim},
  journal={R version},
  volume={90},
  pages={1--66},
  year={2019},
  publisher={The R Foundation Vienna, Austria}
}

@article{friedman2001greedy,
  title={Greedy function approximation: a gradient boosting machine},
  author={Friedman, Jerome H},
  journal={Annals of statistics},
  pages={1189--1232},
  year={2001},
  publisher={JSTOR}
}

@article{greenwell2019package,
  title={Package ‘gbm’},
  author={Greenwell, Brandon and Boehmke, Bradley and Cunningham, Jay and Developers, GBM and Greenwell, Maintainer Brandon},
  journal={R package version},
  volume={2},
  number={5},
  year={2019}
}

@article{ketzler2021climate,
  title={The climate of Norway},
  author={Ketzler, Gunnar and R{\"o}mer, Wolfgang and Beylich, Achim A},
  journal={Landscapes and landforms of Norway},
  pages={7--29},
  year={2021},
  publisher={Springer}
}

@article{hijmans2017package,
  title={Package ‘dismo’},
  author={Hijmans, Robert J and Phillips, Steven and Leathwick, John and Elith, Jane and Hijmans, Maintainer Robert J},
  journal={Circles},
  volume={9},
  number={1},
  pages={1--68},
  year={2017}
}

@article{breiman2001random,
  title={Random forests},
  author={Breiman, Leo},
  journal={Machine learning},
  volume={45},
  pages={5--32},
  year={2001},
  publisher={Springer}
}

@article{dormann2007effects,
  title={Effects of incorporating spatial autocorrelation into the analysis of species distribution data},
  author={Dormann, Carsten F},
  journal={Global ecology and biogeography},
  volume={16},
  number={2},
  pages={129--138},
  year={2007},
  publisher={Wiley Online Library}
}

@article{lennon2000red,
  title={Red-shifts and red herrings in geographical ecology},
  author={Lennon, Jack J},
  journal={Ecography},
  volume={23},
  number={1},
  pages={101--113},
  year={2000},
  publisher={Wiley Online Library}
}

@article{probst2019tunability,
  title={Tunability: Importance of hyperparameters of machine learning algorithms},
  author={Probst, Philipp and Boulesteix, Anne-Laure and Bischl, Bernd},
  journal={The Journal of Machine Learning Research},
  volume={20},
  number={1},
  pages={1934--1965},
  year={2019},
  publisher={JMLR. org}
}

@article{probst2019hyperparameters,
  title={Hyperparameters and tuning strategies for random forest},
  author={Probst, Philipp and Wright, Marvin N and Boulesteix, Anne-Laure},
  journal={Wiley Interdisciplinary Reviews: data mining and knowledge discovery},
  volume={9},
  number={3},
  pages={e1301},
  year={2019},
  publisher={Wiley Online Library}
}

@article{santini2021assessing,
  title={Assessing the reliability of species distribution projections in climate change research},
  author={Santini, Luca and Ben{\'\i}tez-L{\'o}pez, Ana and Maiorano, Luigi and {\v{C}}engi{\'c}, Mirza and Huijbregts, Mark AJ},
  journal={Diversity and Distributions},
  volume={27},
  number={6},
  pages={1035--1050},
  year={2021},
  publisher={Wiley Online Library}
}

@article{svenning2011applications,
  title={Applications of species distribution modeling to paleobiology},
  author={Svenning, Jens-Christian and Fl{\o}jgaard, Camilla and Marske, Katharine A and N{\'o}gues-Bravo, David and Normand, Signe},
  journal={Quaternary Science Reviews},
  volume={30},
  number={21-22},
  pages={2930--2947},
  year={2011},
  publisher={Elsevier}
}

@inproceedings{beery2021species,
  title={Species distribution modeling for machine learning practitioners: A review},
  author={Beery, Sara and Cole, Elijah and Parker, Joseph and Perona, Pietro and Winner, Kevin},
  booktitle={ACM SIGCAS conference on computing and sustainable societies},
  pages={329--348},
  year={2021}
}

@article{pichler2023machine,
  title={Machine learning and deep learning—A review for ecologists},
  author={Pichler, Maximilian and Hartig, Florian},
  journal={Methods in Ecology and Evolution},
  volume={14},
  number={4},
  pages={994--1016},
  year={2023},
  publisher={Wiley Online Library}
}

@article{kodinariya2013review,
  title={Review on determining number of Cluster in K-Means Clustering},
  author={Kodinariya, Trupti M and Makwana, Prashant R and others},
  journal={International Journal},
  volume={1},
  number={6},
  pages={90--95},
  year={2013}
}

@article{bradley1997use,
  title={The use of the area under the ROC curve in the evaluation of machine learning algorithms},
  author={Bradley, Andrew P},
  journal={Pattern recognition},
  volume={30},
  number={7},
  pages={1145--1159},
  year={1997},
  publisher={Elsevier}
}

@article{shabani2018assessing,
  title={Assessing accuracy methods of species distribution models: AUC, specificity, sensitivity and the true skill statistic},
  author={Shabani, Farzin and Kumar, Lalit and Ahmadi, Mohsen},
  journal={Global Journal of Human-Social Science: B Geography, Geo-Sciences, Environmental Science \& Disaster Management},
  volume={18},
  number={1},
  year={2018}
}

@article{willmott2005advantages,
  title={Advantages of the mean absolute error (MAE) over the root mean square error (RMSE) in assessing average model performance},
  author={Willmott, Cort J and Matsuura, Kenji},
  journal={Climate research},
  volume={30},
  number={1},
  pages={79--82},
  year={2005}
}

@article{geisser1975predictive,
  title={The predictive sample reuse method with applications},
  author={Geisser, Seymour},
  journal={Journal of the American statistical Association},
  volume={70},
  number={350},
  pages={320--328},
  year={1975},
  publisher={Taylor \& Francis}
}

@inproceedings{chen2016xgboost,
  title={Xgboost: A scalable tree boosting system},
  author={Chen, Tianqi and Guestrin, Carlos},
  booktitle={Proceedings of the 22nd ACM SIGKDD international conference on knowledge discovery and data mining},
  pages={785--794},
  year={2016}
}

@article{ke2017lightgbm,
  title={Lightgbm: A highly efficient gradient boosting decision tree},
  author={Ke, Guolin and Meng, Qi and Finley, Thomas and Wang, Taifeng and Chen, Wei and Ma, Weidong and Ye, Qiwei and Liu, Tie-Yan},
  journal={Advances in neural information processing systems},
  volume={30},
  year={2017}
}

@article{brenning2005spatial,
  title={Spatial prediction models for landslide hazards: review, comparison and evaluation},
  author={Brenning, A},
  journal={Natural Hazards and Earth System Sciences},
  volume={5},
  number={6},
  pages={853--862},
  year={2005},
  publisher={Copernicus Publications G{\"o}ttingen, Germany}
}

@incollection{getis2009spatial,
  title={Spatial autocorrelation},
  author={Getis, Arthur},
  booktitle={Handbook of applied spatial analysis: Software tools, methods and applications},
  pages={255--278},
  year={2009},
  publisher={Springer}
}

@article{negret2020effects,
  title={Effects of spatial autocorrelation and sampling design on estimates of protected area effectiveness},
  author={Negret, Pablo Jose and Marco, Moreno Di and Sonter, Laura J and Rhodes, Jonathan and Possingham, Hugh P and Maron, Martine},
  journal={Conservation Biology},
  volume={34},
  number={6},
  pages={1452--1462},
  year={2020},
  publisher={Wiley Online Library}
}

@misc{gbif_2022,
  title = {GBIF.org (22 November 2022) GBIF Occurrence Download https://doi.org/10.15468/dl.m7npwp}
}

@article{aiello2015spthin,
  title={spThin: an R package for spatial thinning of species occurrence records for use in ecological niche models},
  author={Aiello-Lammens, Matthew E and Boria, Robert A and Radosavljevic, Aleksandar and Vilela, Bruno and Anderson, Robert P},
  journal={Ecography},
  volume={38},
  number={5},
  pages={541--545},
  year={2015},
  publisher={Wiley Online Library}
}

@Manual{lightgbm,
  title = {lightgbm: Light Gradient Boosting Machine},
  author = {Yu Shi and Guolin Ke and Damien Soukhavong and James Lamb and Qi Meng and Thomas Finley and Taifeng Wang and Wei Chen and Weidong Ma and Qiwei Ye and Tie-Yan Liu and Nikita Titov and David Cortes},
  year = {2024},
  note = {R package version 4.3.0.99},
  url = {https://github.com/Microsoft/LightGBM},
}

@article{adhikari2023global,
  title={Global spatial distribution of Chromolaena odorata habitat under climate change: Random forest modeling of one of the 100 worst invasive alien species},
  author={Adhikari, Pradeep and Lee, Yong Ho and Poudel, Anil and Hong, Sun Hee and Park, Yong-Soon},
  journal={Scientific Reports},
  volume={13},
  number={1},
  pages={9745},
  year={2023},
  publisher={Nature Publishing Group UK London}
}

@article{maglietta2023environmental,
  title={Environmental variables and machine learning models to predict cetacean abundance in the Central-eastern Mediterranean Sea},
  author={Maglietta, Rosalia and Saccotelli, Leonardo and Fanizza, Carmelo and Telesca, Vito and Dimauro, Giovanni and Causio, Salvatore and Lecci, Rita and Federico, Ivan and Coppini, Giovanni and Cipriano, Giulia and others},
  journal={Scientific Reports},
  volume={13},
  number={1},
  pages={2600},
  year={2023},
  publisher={Nature Publishing Group UK London}
}

@article{hijmans2015package,
  title={Package ‘raster’},
  author={Hijmans, Robert J and Van Etten, Jacob and Cheng, Joe and Mattiuzzi, Matteo and Sumner, Michael and Greenberg, Jonathan A and Lamigueiro, Oscar Perpinan and Bevan, Andrew and Racine, Etienne B and Shortridge, Ashton and others},
  journal={R package},
  volume={734},
  pages={473},
  year={2015}
}

@article{bivand2015package,
  title={Package ‘rgdal’},
  author={Bivand, Roger and Keitt, Tim and Rowlingson, Barry and Pebesma, Edzer and Sumner, Michael and Hijmans, Robert and Rouault, Even and Bivand, Maintainer Roger},
  journal={Bindings for the Geospatial Data Abstraction Library. Available online: https://cran. r-project. org/web/packages/rgdal/index. html (accessed on 15 October 2017)},
  volume={172},
  year={2015}
}

@article{pebesma2018simple,
  title={Simple features for R: standardized support for spatial vector data.},
  author={Pebesma, Edzer J and others},
  journal={R J.},
  volume={10},
  number={1},
  pages={439},
  year={2018}
}

@article{hijmans2022package,
  title={Package ‘terra’},
  author={Hijmans, Robert J and Bivand, Roger and Forner, Karl and Ooms, Jeroen and Pebesma, Edzer and Sumner, Michael D},
  journal={Maintainer: Vienna, Austria},
  year={2022}
}

@article{kuhn2020package,
  title={Package ‘caret’},
  author={Kuhn, Max and Wing, Jed and Weston, Steve and Williams, Andre and Keefer, Chris and Engelhardt, Allan and Cooper, Tony and Mayer, Zachary and Kenkel, Brenton and Team, R Core and others},
  journal={The R Journal},
  volume={223},
  number={7},
  year={2020}
}

@article{koldasbayeva2024challenges,
  title={Challenges in data-driven geospatial modeling for environmental research and practice},
  author={Koldasbayeva, Diana and Tregubova, Polina and Gasanov, Mikhail and Zaytsev, Alexey and Petrovskaia, Anna and Burnaev, Evgeny},
  journal={Nature Communications},
  volume={15},
  number={1},
  pages={10700},
  year={2024},
  publisher={Nature Publishing Group UK London}
}

@article{hijmans2012cross,
  title={Cross-validation of species distribution models: removing spatial sorting bias and calibration with a null model},
  author={Hijmans, Robert J},
  journal={Ecology},
  volume={93},
  number={3},
  pages={679--688},
  year={2012},
  publisher={Wiley Online Library}
}

@techreport{bradburn2011,
  author    = {Mark James Bradburn and Aimee A. Keller and Beth Helene Horness},
  title     = {The 2003 to 2008 U.S. West Coast bottom trawl surveys of groundfish resources off Washington, Oregon, and California: Estimates of distribution, abundance, length, and age composition},
  year      = {2011},
  institution = {U.S. Department of Commerce, NOAA Technical Memorandum NMFS-NWFSC-114},
  number    = {NMFS-NWFSC-114},
  pages     = {323},
  type      = {NOAA Technical Memorandum}
}

@article{radosavljevic2014making,
  title={Making better Maxent models of species distributions: complexity, overfitting and evaluation},
  author={Radosavljevic, Aleksandar and Anderson, Robert P},
  journal={Journal of biogeography},
  volume={41},
  number={4},
  pages={629--643},
  year={2014},
  publisher={Wiley Online Library}
}

@article{syfert2013effects,
  title={The effects of sampling bias and model complexity on the predictive performance of MaxEnt species distribution models},
  author={Syfert, Mindy M and Smith, Matthew J and Coomes, David A},
  journal={PloS one},
  volume={8},
  number={2},
  pages={e55158},
  year={2013},
  publisher={Public Library of Science San Francisco, USA}
}

@article{vignali2020sdmtune,
  title={SDMtune: An R package to tune and evaluate species distribution models},
  author={Vignali, Sergio and Barras, Arnaud G and Arlettaz, Rapha{\"e}l and Braunisch, Veronika},
  journal={Ecology and Evolution},
  volume={10},
  number={20},
  pages={11488--11506},
  year={2020},
  publisher={Wiley Online Library}
}

@article{velazco2022flexsdm,
  title={flexsdm: An r package for supporting a comprehensive and flexible species distribution modelling workflow},
  author={Velazco, Santiago Jos{\'e} El{\'\i}as and Rose, Miranda Brooke and de Andrade, Andr{\'e} Felipe Alves and Minoli, Ignacio and Franklin, Janet},
  journal={Methods in Ecology and Evolution},
  volume={13},
  number={8},
  pages={1661--1669},
  year={2022},
  publisher={Wiley Online Library}
}

@article{steen2021spatial,
  title={Spatial thinning and class balancing: Key choices lead to variation in the performance of species distribution models with citizen science data},
  author={Steen, Valerie A and Tingley, Morgan W and Paton, Peter WC and Elphick, Chris S},
  journal={Methods in Ecology and Evolution},
  volume={12},
  number={2},
  pages={216--226},
  year={2021},
  publisher={Wiley Online Library}
}

@article{BORIA201473,
title = {Spatial filtering to reduce sampling bias can improve the performance of ecological niche models},
journal = {Ecological Modelling},
volume = {275},
pages = {73-77},
year = {2014},
issn = {0304-3800},
doi = {https://doi.org/10.1016/j.ecolmodel.2013.12.012},
author = {Robert A. Boria and Link E. Olson and Steven M. Goodman and Robert P. Anderson}}

@article{hirzel2006evaluating,
  title={Evaluating the ability of habitat suitability models to predict species presences},
  author={Hirzel, Alexandre H and Le Lay, Gwena{\"e}lle and Helfer, V{\'e}ronique and Randin, Christophe and Guisan, Antoine},
  journal={Ecological modelling},
  volume={199},
  number={2},
  pages={142--152},
  year={2006},
  publisher={Elsevier}
}

@article{mila2022nearest,
  title={Nearest neighbour distance matching Leave-One-Out Cross-Validation for map validation},
  author={Mila, Carles and Mateu, Jorge and Pebesma, Edzer and Meyer, Hanna},
  journal={Methods in Ecology and Evolution},
  volume={13},
  number={6},
  pages={1304--1316},
  year={2022},
  publisher={Wiley Online Library}
}

@book{hyndman2018forecasting,
  title={Forecasting: principles and practice},
  author={Hyndman, Rob J and Athanasopoulos, George},
  year={2018},
  publisher={OTexts}
}

@article{reynolds2007daily,
  title={Daily high-resolution-blended analyses for sea surface temperature},
  author={Reynolds, Richard W and Smith, Thomas M and Liu, Chunying and Chelton, Dudley B and Casey, Kenneth S and Schlax, Michael G},
  journal={Journal of climate},
  volume={20},
  number={22},
  pages={5473--5496},
  year={2007}
}

@article{lobo2008auc,
  title={AUC: a misleading measure of the performance of predictive distribution models},
  author={Lobo, Jorge M and Jim{\'e}nez-Valverde, Alberto and Real, Raimundo},
  journal={Global ecology and Biogeography},
  volume={17},
  number={2},
  pages={145--151},
  year={2008},
  publisher={Wiley Online Library}
}

@article{bracho2024spatially,
  title={Spatially explicit metrics improve the evaluation of species distribution models facing sampling biases},
  author={Bracho-Est{\'e}vanez, Claudio A and Arenas-Castro, Salvador and Gonz{\'a}lez-Varo, Juan P and Gonz{\'a}lez-Moreno, Pablo},
  journal={Ecological Informatics},
  volume={84},
  pages={102916},
  year={2024},
  publisher={Elsevier}
}

@article{ward2015using,
  title={Using spatiotemporal species distribution models to identify temporally evolving hotspots of species co-occurrence},
  author={Ward, Eric J and Jannot, Jason E and Lee, Yong-Woo and Ono, Kotaro and Shelton, Andrew O and Thorson, James T},
  journal={Ecological Applications},
  volume={25},
  number={8},
  pages={2198--2209},
  year={2015},
  publisher={Wiley Online Library}
}





\end{document}